\documentclass[11pt,twoside]{article}
\usepackage{asp2006}
\usepackage{epsf}
\usepackage{psfig}
\usepackage{lscape}

\begin{document}
\title{A Multifrequency Study of Double-Double Radio Galaxies}
\author{M. Jamrozy,$^1$ C. Konar,$^2$ D.J. Saikia,$^3$ and J. Machalski$^1$}
\affil{$^1$Jagiellonian University, 30--244 Krak\'ow, Poland  \\
$^2$IUCAA, Pune University Campus, Pune 411007, India \\
$^3$National Centre for Radio Astrophysics, TIFR, Pune 411007, India}

\begin{abstract}
One of the striking examples of episodic activity in active galactic nuclei are the double-double 
radio galaxies (DDRGs) with two pairs of oppositely-directed radio lobes from two different cycles 
of activity. We illustrate, using the DDRG J1453+3308 as an example,  that observations over a
wide range of frequencies using both the GMRT 
and the VLA can be used to determine the spectra of the inner and outer lobes, estimate their spectral 
ages, estimate the time scales of episodic activity, and examine any difference in the injection spectra 
in the two cycles of activity. Low-frequency GMRT observations also suggest that DDRGs and triple-double
radio galaxies are rather rare.  
\end{abstract}

\section{Introduction}
For active galactic nuclei (AGN), an important and interesting issue is the duration of their active phase 
and the time scales for recurrence of such periods of activity. It is believed widely that the central 
activity is related to the feeding of a supermassive black hole in the centre of the galaxy. 
One of the more striking examples of episodic activity is when a new pair of 
radio lobes is seen closer to the nucleus in a radio loud AGN 
before the older and more distant radio lobes have faded. Such sources 
have been named double-double radio galaxies (DDRGs; Schoenmakers et al. 2000). A DDRG consists of a pair of 
double radio sources with a common core, where the two lobes of the inner double have an edge-brightened
radio morphology. In such sources the newly-formed jets  may propagate outwards through the cocoon formed 
by the earlier cycle of activity rather than the general intergalactic or intracluster medium. In Table~1 
we summarise the presently known large-scale DDRGs, including a couple of candidates. All of these are 
identified with galaxies (not quasars) and the linear sizes of the outer doubles of most DDRGs are 
larger than $\sim$1 Mpc. However, in addition to the DDRGs there are other signatures of episodic activity.  

For example, the radio galaxy 4C29.30 (J0840+2949), which is associated with
a bright (R $\rm\sim 15^{m}$) host elliptical galaxy 
at a redshift of 0.06471$\pm$0.00013, has an 
inner double-lobed radio source with two prominent
hotspots separated by 29 arcsec (36 kpc), embedded in a large-scale halo which 
has an angular size of $\sim$520 arcsec (639 kpc). 
The radio luminosity of the inner double at 1400 MHz is
5.5$\times$10$^{24}$ W Hz$^{-1}$,  which is significantly below the dividing line of the Fanaroff-Riley
classes, while that of the entire source is 7.4$\times$10$^{24}$ W Hz$^{-1}$. It is interesting to
note that in some of the DDRGs, the luminosity of the inner double is in the FRI category although its 
structure resembles that of FRII radio sources (cf. Saikia et al. 2006).
The spectral age of the inner double is estimated to be
less than $\sim$33 Myr. The extended diffuse emission has a steep spectrum with a spectral
index of $\sim$1.3 and a break frequency less than $\sim$240 MHz. The spectral age is
greater than $\sim$200 Myr, suggesting that the extended diffuse emission is due to 
an earlier cycle of activity (Jamrozy et al. 2007). 

In addition, diffuse relic radio emission due to an earlier cycle of activity may
also be visibile around radio sources which may not be characterised by a `classical
double' structure with hotspots at the outer edges. Examples of such sources include
Her A (Gizani \& Leahy 2003), 3CR 310 (e.g. Leahy, Pooley, \& Riley 1986), and Cen A (e.g. 
Morganti et al. 1999).  Giacintucci et al. (2007) have explored the
possibility that the interesting wide-angle tailed source at the centre of Abell 2372 might be an 
example of a restarted radio galaxy.

\begin{center}
\begin{table}[t]
\caption{The sample of known DDRGs}
\label{ddrg_src}
{\footnotesize
\begin{tabular}{l c r r r r r r r r}

\hline
Source name &Opt.& z &$\rm d_{in}$& $\rm \theta_{in}$   &$\rm d_{out}$&$\rm \theta_{out}$ &$\rm S_{in}$ &$\rm S_{out}$&Ref.  \\
            &Id. &   & kpc        &$\prime\prime$ &  kpc        &$\prime\prime$&mJy          & mJy         &      \\
    (1)     &(2) &(3)& (4)        & (5)           &  (6)        &  (7)         &  (8)        &  (9)        & (10) \\
\hline
J0041$+$3224&G&0.45 & 171     & 30     & 969      &170    &525     &409      & 1       \\
J0116$-$4722&G&0.15 & 460     & 180    & 1447     &570    &260     &2640     & 2       \\
J0921$+$4538&G&0.17 &  69     & 24     & 433      &150    &90      &8046     & 3, 4, 5 \\
J0927$+$3510&G&0.55 & 575     & 90     & 2206     &345    &~3      &96       & 6       \\
J0929$+$4146&G&0.37 & 652     & 130    & 1875     &372    &64      &99       & 7, 8, (*)\\
J1006$+$3454&G&0.10 & 1.7     & 1      & 4249     &2310   &2500    &3300     & 9, 10, 11\\
J1158$+$2621&G&0.11 & 138     & 69     & 483      &240    &67      &962      & 12      \\
J1242$+$3838&G&0.30 & 251     & 57     & 602      &136    &8       &24       & 7       \\
J1247$+$6723&G&0.11 &0.014    & 0.01   & 1195     &618    &260     &126      & 13, 14  \\
J1453$+$3308&G&0.25 & 159     & 41     & 1297     &336    &34      &426      & 7, 15   \\
J1548$-$3216&G&0.11 & 313     & 160    & 961      &493    &78      &1722     & 16, 17  \\
J1604$+$3438&G&0.28 & 212     & 50     & 845      &200    &~5      &146      & 6       \\
J1835$+$6204&G&0.52 & 369     & 60     & 1379     &222    &200     &604      & 7       \\
J2223$-$0206&G&0.06 & 130     & 121    & 612      &570    &55      &5260     & 18, 19  \\
\hline
\end{tabular}

Column~1: source name; columns~2 and 3: optical identification and redshift; 
columns~4 and 6: projected linear sizes of the inner and outer lobes in kpc (calculated for 
$\rm H_{0}$=71 km s$^{-1}$ Mpc$^{-1}$, $\rm \Omega_{M}$ = 0.27, and $\rm \Omega_{vac}$ = 0.73); 
columns~5 and 7: largest angular size of the inner and outer lobes in arcsec; columns~8 and 9: 1400-MHz 
flux density of the inner and outer lobes in mJy; column~10: references.  
1: Saikia et al. (2006); 2: Saripalli, Subrahmanyan, \& Udaya Shankar (2002); 
3: Perley et al. (1980); 4: Bridle, Perley, \& Henriksen (1986); 5: Clarke et al. (1992); 
6: Machalski et al. (2006); 7: Schoenmakers et al. (2000); 8: Brocksopp et al. (2007);
9: Willis, Strom, \& Wilson (1974); 10: Strom \& Willis (1980); 11: Schilizzi et al. (2001); 
12: Owen \& Ledlow (1997); 13: Marecki et al. (2003); 14: Bondi et al. (2004); 
15: Konar et al. (2006); 16: Saripalli, Subrahmanyan, \& Udaya Shankar (2003);
17: Safouris et al. (2008); 18: Kronberg, Wielebinski, \& Graham (1986); 19: Leahy et al. (1997); 
(*) a triple-double radio galaxy.

}
\end{table}
\end{center}

\section{A Multifrequency Study of J1453+3308}
The lobes of the outer double of J1453+3308 are separated by 336 arcsec (1297 kpc), and 
their radio luminosity is above the FRI/FRII break although they do not show any prominent 
compact hot-spots. The inner lobes have a separation of 41 arcsec (159 kpc), 
and their luminosity is below the dividing FRI/FRII line although they have an 
edge-brightened structure (Figure 1). 

\begin{figure}[t]
\hspace{0.9cm}
\hbox{
  \psfig{file=J1453+3309_610.PS,width=2.0in,angle=0}
            \vbox{
                 \psfig{file=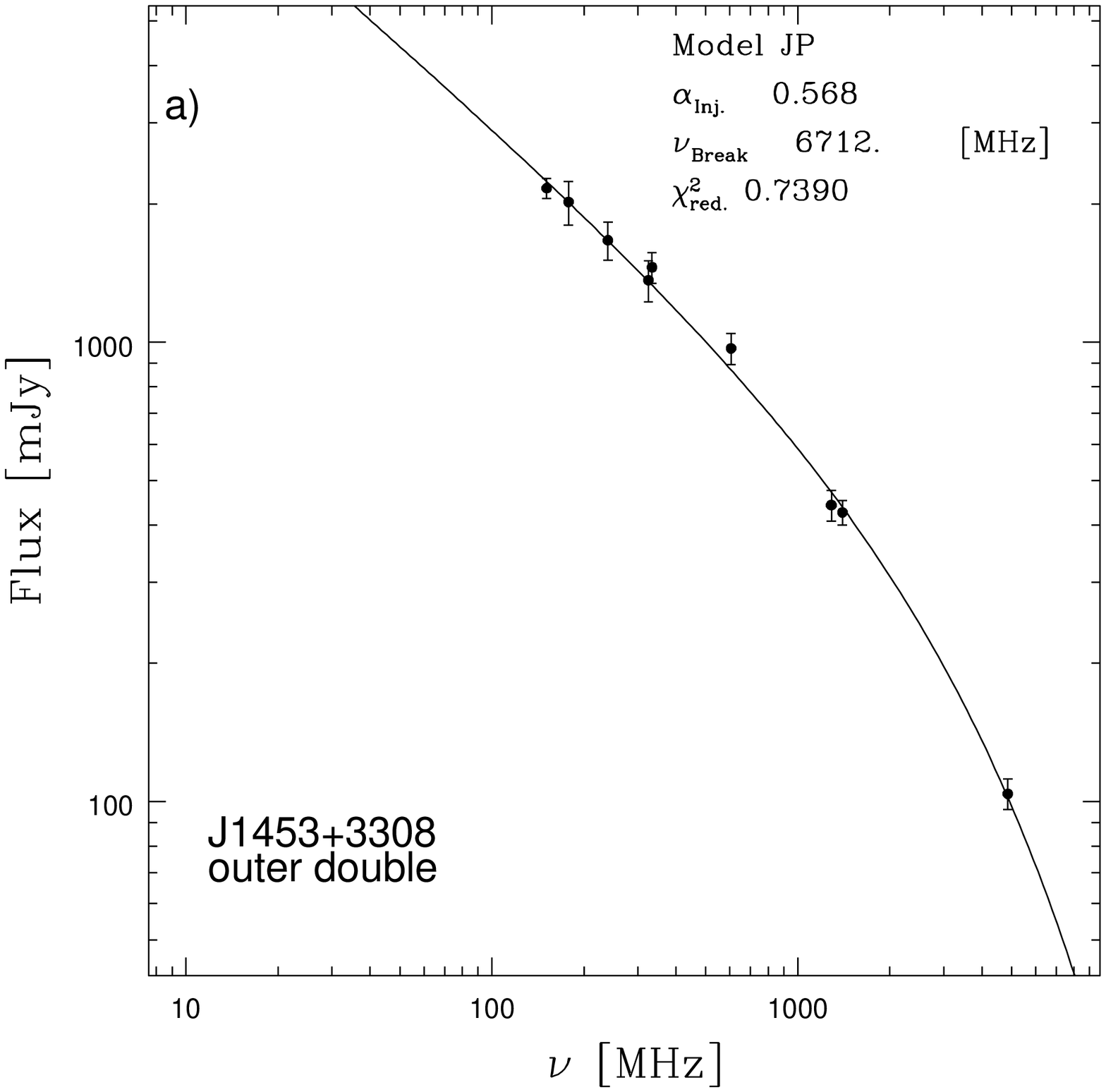,width=2.0in,angle=0}
                 \psfig{file=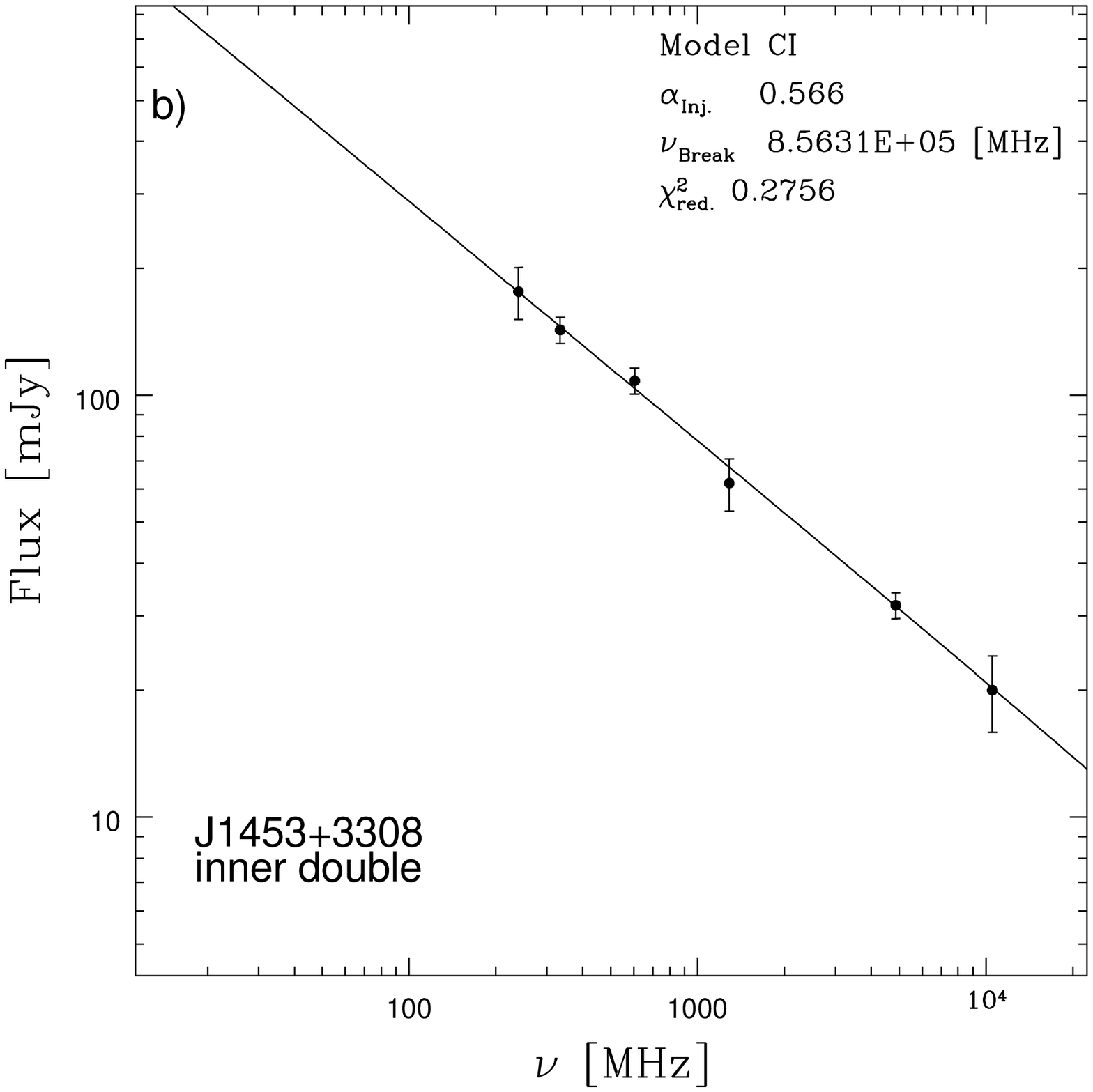,width=2.0in,angle=0}
                 }
     }
\caption[]{Left: The GMRT image of the DDRG J1453+3308 at 605 MHz with an angular resolution of $\sim$5.4 arcsec.
Spectra of the outer and inner doubles fitted with the models of radiative
losses. Upper right panel: the outer double fitted with the Jaffe-Perola model.
Lower right panel: the inner double fitted with the continuous-injection model (Konar et al. 2006).
}
\end{figure}

Spectral ages have been determined using the standard
theory describing the time-evolution of the emission spectrum from particles with an
initial power-law energy distribution, corresponding to an injection 
spectral index $\alpha_{\rm inj}$ (Murgia 1996). The spectral break frequency above which the radio spectrum
steepens from the injected power law, $\nu_{\rm Break}$, is related to the spectral age. 
In order to determine a value of $\alpha_{\rm inj}$, we fitted the CI 
(continuous injection) and JP (Jaffe \& Perola 1973) models of radiative losses 
to the flux densities of the inner and outer lobes, respectively. The fits of the models to the flux densities of 
both the inner lobes as well as the entire outer double are shown in  Figure 1. 
It is worth noting that both the fitted values of 
$\alpha_{\rm inj}$ are $\sim0.57$. The spectral age obtained for the inner double is of $\la$2 Myr, while
the maximum ages for the northern 
and southern lobes are $\sim$47 Myr and $\sim$58 Myr, respectively. 
This indicates a mean separation velocity of the 
lobe's head from the radio-emitting plasma of 0.036c. However, assuming presence of a backflow with a
backward speed comparable to the advance speed of the head, an average advance speed would be
about 0.018c which gives a maximum age of $\sim$130 Myr (Konar et al. 2006 for more details).

\begin{figure}[t]
\hspace{2cm}
\vbox{
  \psfig{file=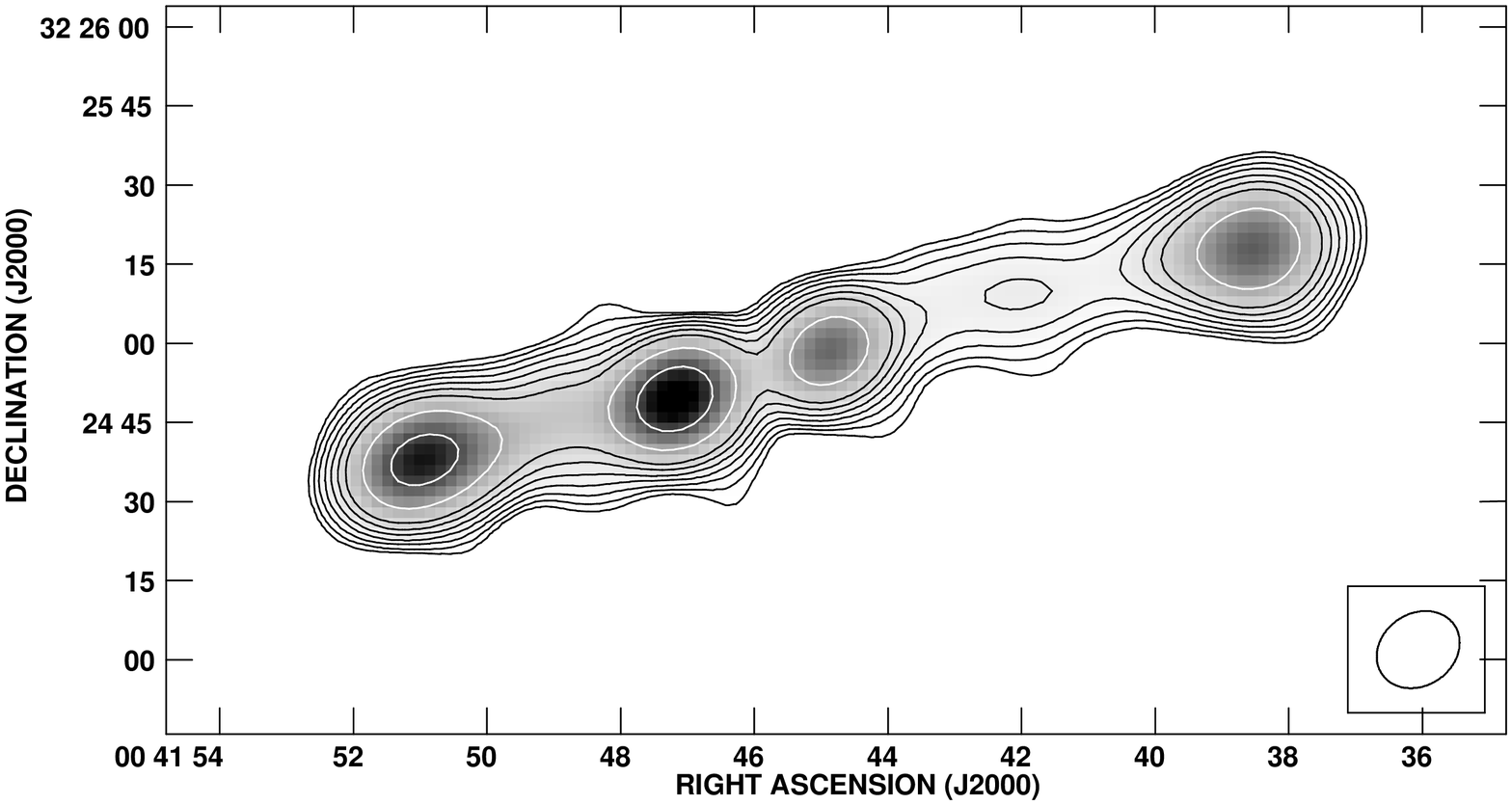,width=3.5in,angle=0}
            \hbox{
                 \psfig{file=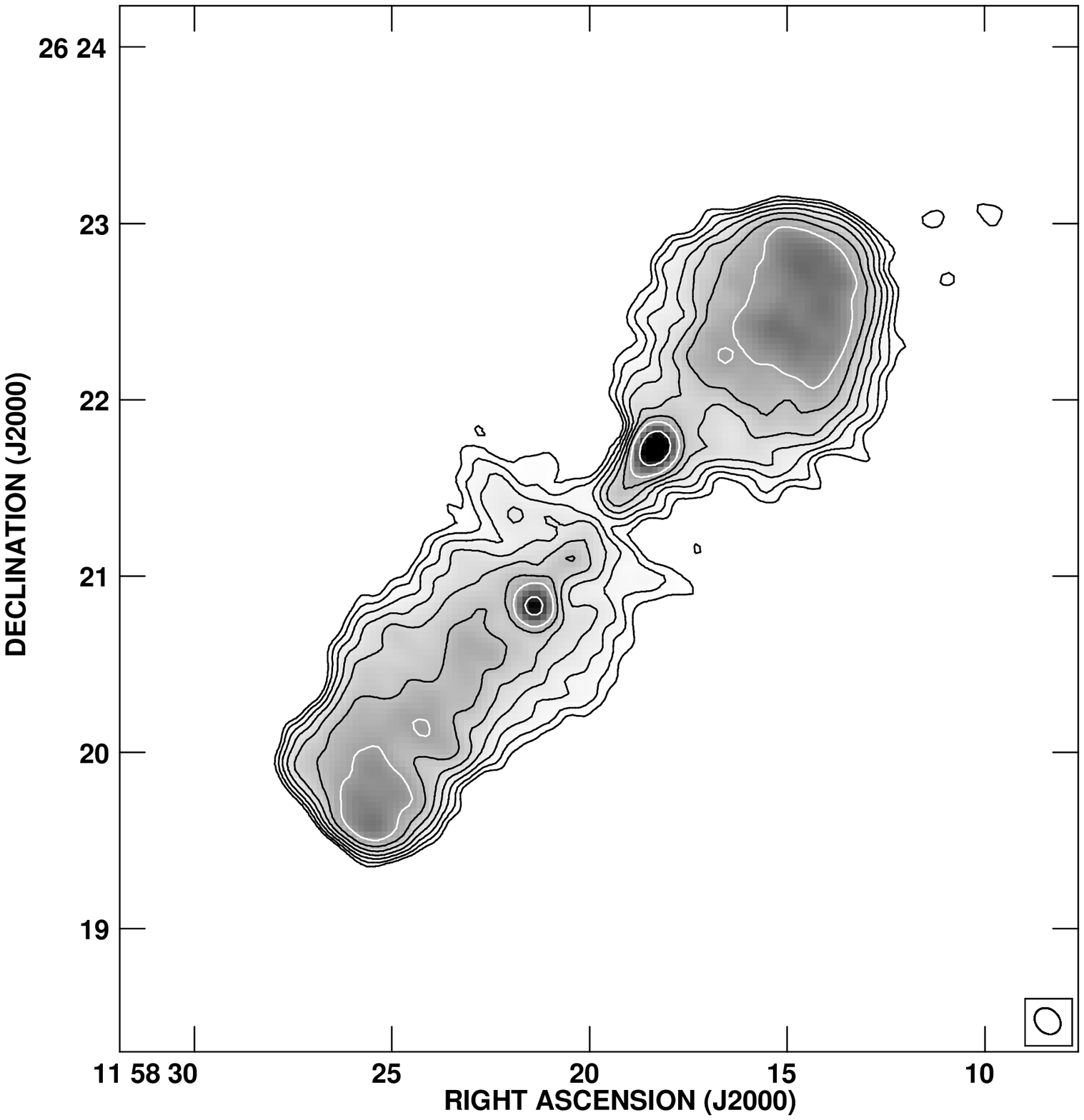,width=2.0in,angle=0}
                 \psfig{file=J1548_3DS5.FINAL.A.PS,width=1.7in,angle=0}
                 }
     }
\caption[]{The GMRT images of the DDRGs J0041+3224 at 240 MHz (upper panel), and J1158$+$2621 (lower left panel)
and J1548$-$3216 (lower right panel) at $\sim$333 MHz. }
\end{figure}

\section{Discussion and Concluding Remarks}
It is interesting to note that only a small number of radio galaxies appear to 
exhibit clear signatures of episodic activity (see Table 1). A deep low-frequency
search with the GMRT for evidence of fossil radio lobes which could be due to an earlier cycle of 
episodic activity showed no unambiguous examples in 
a sample of 374 sources (Sirothia et al. 2009). Most of the sources in this sample are however small,
consistent with the trend that radio galaxies with evidence of episodic activity tend
to be large, often over a Mpc in size. This suggests that the time scales of episodic 
activity are in the range of 10$^6$ to 10$^8$ yr.  

In the list of DDRGs (Table 1), there is only one case of a triple-double radio
galaxy (Brocksopp et al. 2007), and so far no case of a DDRG associated with a quasar has
been reported. Our low-frequency images of known DDRGs to search for an even earlier cycle
of episodic activity (Figure 2) have not shown any candidate triple-double radio galaxy.

Both J1453+3308 and 4C29.30 appear to have similar injection spectral indices for both
the outer and inner doubles, although they are traversing through very different environments.
A similar trend has also been noted for PKS 1545$-$321 (Safouris et al. 2008). 
If the episodic activity is triggered by a fresh supply of gas one might find evidence of cold gas. 
Saikia, Gupta, \& Konar (2007) reported the detection of H{\sc i} in absorption towards the 
DDRG J1247+6723 and suggested a strong
relationship between episodic activity and detection of H{\sc i} gas with complex line
profiles.  Konar et al. (2006) and Jamrozy et al. (2007) reported 
evidence of variability in the weak radio cores of J1453+3308 and 4C29.30. 
All these aspects need to be explored for a larger sample of radio sources with
signs of episodic activity, to put better constraints on models of these objects. 

\acknowledgements M.J. would like to thank IUCAA for their warm hospitality and support 
during his visit to Pune at about the time of the the LFRU conference period.

\end{document}